\documentclass[preprint]{revtex4}%
\usepackage{ae,graphics,epsfig,graphicx}
\usepackage{dcolumn}
\usepackage{bm}
\usepackage{amsmath}
\usepackage{amsfonts}
\usepackage{amssymb}
\usepackage{graphicx}%
\providecommand{\U}[1]{\protect\rule{.1in}{.1in}}
\begin{document}
\title{Method for calculating the electronic structure of correlated materials from a
truly first-principles LDA+U scheme \textbf{ }}
\author{K. Karlsson$^{1,2}$, F. Aryasetiawan$^{3,4}$ and O. Jepsen$^{2}$,}
\affiliation{$^{1}$\textit{Department of Life Sciences, H\"{o}gskolan i Sk\"{o}vde, 54128
Sk\"{o}vde, Sweden,}}
\affiliation{$^{2}$\textit{Max Planck Institut f\"{u}r Festk\"{o}rperforschung, D-705 06
Stuttgart, Germany,}\emph{Japan Science and Technology Agency, CREST,
Kawaguchi 332-0012 Japan}}
\affiliation{$^{3}$\textit{Graduate School of Advanced Integration Science, Chiba
University, Chiba, 263-8522 Japan,}}
\affiliation{$^{4}$\textit{Japan Science and Technology Agency, CREST, Kawaguchi 332-0012
Japan.} }
\date{}

\begin{abstract}
We present a method for calculating the electronic structure of correlated
materials based on a truly \emph{first-principles} LDA+$U$ scheme. Recently we
suggested how to calculate $U$ from first-principles, using a method which we
named constrained RPA (cRPA). The input is simply the Kohn-Sham eigenfunctions
and eigenvalues obtained within the LDA. In our proposed self-consistent
LDA+$U$ scheme, we calculate the LDA+$U$ eigenfunctions and eigenvalues and
use these to extract $U$. The updated $U$ is then used in the next iteration
to obtain a new set of eigenfunctions and eigenvalues and the iteration is
continued until convergence is achieved. The most significant result is that
our numerical approach is indeed stable: it is possible to find the effective
exchange and correlation interaction matrix in a \emph{self-consistent} way,
resulting in a significant improvement over the LDA results, regarding both
the bandgap in $NiO$ and the $f$-band exchange spin-splitting in $Gd$, but
some discrepancies still remain.

\end{abstract}
\maketitle

\section{Introduction}

The interest for a fundamental understanding of strongly correlated systems
has led to the development of a number of electronic structure methods. Among
the most successful are the LDA+$U$ approach proposed by Anisimov and
coworkers \cite{zaan} and the dynamical mean-field theory (DMFT) proposed by
Georges and coworkers \cite{rev2,pru}. Very recently a new scheme, dubbed the
LDA+Gutzwiller method \cite{deng}, for treating strong electron correlations
was introduced. In all these methods the strong Coulomb onsite correlations
for electrons residing in the localized orbitals are explicitly taken care of
via a set of Hubbard like \textit{parameters }or the Hubbard \emph{U}. This is
evidently unsatisfactory from the point of view of quantitative prediction of
materials properties, since optical and magnetic excitations are of vital
importance in many technological applications such as solar cell design,
optical memories, photo luminescent devices (semiconductor lasers and diodes)
and photo chemical reactions. Often it has been shown that by adjusting the
Hubbard $U$ one can get results in good agreement with experiment but not for
a good reason. Hereby lies the importance of determining $U$ entirely from first-principles.

Over the last two decades a number of methods for calculating the Hubbard
\emph{U} from first principles have been proposed. The pioneering work may be
traced back to the paper by Gunnarsson \textit{et al.} \cite{gunnar} who
proposed to calculate \emph{U} using the constrained LDA (cLDA) scheme. A few
years ago, a new method for calculating the Hubbard \emph{U}, named the
constrained random-phase approximation (cRPA) method in analogy to the
cLDA\ method, was proposed \cite{ferdi2}. The method allows for a systematic
and precise determination of the Hubbard \emph{U} entirely from
first-principles from the knowledge of the bandstructure alone. The method was
based on the intuitive idea that the Hubbard \emph{U} can be viewed as a
Coulomb interaction screened by the polarization of the whole system excluding
the polarization arising from a set of bands which are treated in the Hubbard
model. In other words, the Hubbard \emph{U} when further screened by the
electrons in the Hubbard model yields the screened interaction of the full
system. This intuitive idea was recently shown to be rigorously correct and
the cRPA is just an approximate way of calculating the screened interaction
\emph{U }within the random-phase approximation (RPA).

By determining the Hubbard \emph{U} from first-principles the cRPA method
offers the possibility of making methods based on the Hubbard \emph{U} fully
first-principles schemes. The purpose of the present work is to develop a
scheme for calculating the electronic structure of correlated materials based
on a truly \emph{self-consistent} \emph{first-principles} LDA+$U$ scheme. In
conventional LDA+$U$ scheme as it was originally proposed \cite{zaan}, the
Hubbard $U$ is taken as an adjustable parameter which is fixed for a given
calculation. In our proposed self-consistent LDA+$U$ scheme, we calculate the
LDA+$U$ eigenfunctions and eigenvalues and use these to calculate $U$ using
the cRPA method. The new $U$ is then used in the next iteration to obtain a
new set of eigenfunctions and eigenvalues and the iteration is continued until
convergence is achieved. Thus, $U$ is no longer an arbitrarily adjusted
parameter like in the original LDA+$U$ scheme but rather it is determined
self-consistently within the theoretical scheme. Our first target will be to
calculate the electronic structure of the transition metal oxide series and as
a test case we consider NiO, which is regarded as the epitome of Mott-Hubbard
insulators. We are also aiming at obtaining a more satisfactory description of
the electronic structure of the $4f$ electron series which is highly
problematic for the LDA. The path is then opened for more complex materials,
such as magnetic semiconductors, for which no realistic methods are in
existent at present.

In the present paper we present some results for NiO and Gd, which we believe
should provide us with a stringent test of the applicability of our method.
The most important finding is that our numerical approach is indeed stable i.e
it is possible to find $U$ and $J$ self-consistently. The bandgap in NiO and
the spin-splitting of the $f$-bands in Gd are found to compare well with
experiment using our self-consistent determined values of the correlation parameters.

\section{Theory}

\subsection{Constrained RPA}

We first give a short summary of the cRPA method presented in detail elsewhere
\cite{ferdi1,ferdi2}. The fully screened Coulomb interaction is given by%

\begin{equation}
W=[1-vP]^{-1}v
\end{equation}
where \emph{v} is the bare Coulomb interaction and \emph{P }is the
non-interacting polarization given by
\begin{align}
P(\mathbf{r,r}^{\prime};\omega)  &  =\sum_{i}^{occ}\sum_{j}^{unocc}\psi
_{i}(\mathbf{r)}\psi_{i}^{\ast}(\mathbf{r}^{\prime})\psi_{j}^{\ast
}(\mathbf{r)}\psi_{j}(\mathbf{r}^{\prime})\nonumber\\
&  \times\left\{  \frac{1}{\omega-\varepsilon_{j}+\varepsilon_{i}+i0^{+}%
}-\frac{1}{\omega+\varepsilon_{j}-\varepsilon_{i}-i0^{+}}\right\}  . \label{P}%
\end{align}
where $\{\psi_{i},\varepsilon_{i}\}$ are one-particle Bloch eigenfunctions and
eigenvalues corresponding to the system's band structure. For systems with a
narrow $3d$ or $4f$ band crossing the Fermi level, typical of strongly
correlated materials, we may divide the polarization into $P=P_{d}+P_{r}$, in
which $P_{d}$ includes merely the transitions within the narrow band
($3d$-$3d$ \emph{or} $4f$-$4f$ transitions) and $P_{r}$ be the rest of the
polarization, which includes transitions from the 3d band to the rest of the
bands and vice-versa. It was noticed that the following quantity can be
interpreted as the effective interaction among electrons living in the narrow
band (Hubbard \emph{U)}:
\begin{equation}
U(\omega)=[1-vP_{r}(\omega)]^{-1}v \label{Wr}%
\end{equation}
where $U$ can be related to the fully screened interaction \emph{W} by the
following \emph{identity:}
\begin{equation}
W=[1-UP_{d}]^{-1}U. \label{W}%
\end{equation}
This identity explicitly shows that the interaction between the $3d$ or $4f$
electrons is given by a frequency-dependent interaction $U$. Thus the
remaining screening channels in the Hubbard model associated with the
localized $d$ electrons, represented by the $d$-$d$ polarization $P_{d},$
further screen $U$ to give the fully screened interaction $W$. We refer the
method of calculating the Hubbard \emph{U} according to (\ref{Wr}) as cRPA
because we have constrained the polarization to exclude transitions within the
narrow band ($d$-$d$ transitions). Although the formula in (\ref{Wr}) has been
obtained within the RPA, the result is actually exact provided $P_{r}$ is
exact, as was shown recently \cite{ferdi09}.

In the following, we retain only the local components of the effective
interaction on the same atomic site by taking the following matrix element:%

\begin{equation}
U_{L_{1}L_{2},L_{3}L_{4}}=\int d^{3}rd^{3}r^{\prime}\ \phi_{L_{1}}^{\ast
}(\mathbf{r)}\phi_{L_{2}}(\mathbf{r)}U(\mathbf{r,r}^{\prime})\phi_{L_{3}%
}^{\ast}(\mathbf{r}^{\prime}\mathbf{)}\phi_{L_{4}}(\mathbf{r}^{\prime
}\mathbf{)} \label{U}%
\end{equation}
where $\phi_{\zeta}$ is a ${\zeta}$ LMTO \cite{andersen-lmto} orbital ($3d$ or
$4f$) centered on an atomic site and the interaction $U(\mathbf{r,r}^{\prime
})$ is the static ($\omega=0)$ value of Eq. (\ref{Wr}). In calculating
\emph{U} we have approximated $\phi_{\zeta}$ by the "head" of the LMTO, i.e.,
the solution to the Schr\"{o}dinger equation inside the atomic sphere. This is
expected to be a reasonable approximation because the $\zeta$ states are
rather localized. LMTO is just one possible choice for the one-particle
orbitals but other choices are perfectly legitimate. For example, the newly
developed NMTO (where N is the number of energies chosen to span the region of
interest) \cite{nmto} and the recently proposed maximally localized Wannier
orbitals \cite{vanderbilt} are possible choices. It is worth noting that the
\emph{U} entering the Hubbard model will inevitably depend on the choice of
the one-particle basis $\phi_{\zeta}$ defining the annihilation and creation
operators, no matter what method we use to calculate $U(\mathbf{r,r}^{\prime
})$, which is independent of the basis functions used in the band structure method.

\subsection{LDA+U}

In the spirit of the LDA+$U$ approach\cite{zaan}, we introduce an
orbital-dependent exchange-correlation operator
\[
{\large \hat{V}}_{\sigma}=\sum_{RL,R^{\prime}L^{\prime}}|\phi_{RL\sigma
}\rangle V_{RL,R^{\prime}L^{\prime}}^{\sigma}\langle\phi_{R^{\prime}L^{\prime
}\sigma}|
\]
acting among a localized set of electrons. The LMTO head is in general denoted
by site index $R$, angular quantum number $L=(lm)$ and spin $\sigma$. In
addition to the usual single-particle LDA Hamiltonian, we include appropriate
matrix-elements of ${\large \hat{V}}_{\sigma}$. In the
TB-representation\cite{jepsen-TB} we get
\[
\langle\chi_{RL\sigma}^{\mathbf{k}}|{\large \hat{V}}_{\sigma}|\chi_{R^{\prime
}L^{\prime}\sigma}^{\mathbf{k}}\rangle=\sum_{R^{\prime\prime}L^{\prime\prime}%
}\langle\chi_{RL\sigma}^{\mathbf{k}}|\phi_{R^{\prime\prime}L^{\prime\prime
}\sigma}\rangle V_{R^{\prime\prime}L^{\prime\prime}}^{\sigma}\langle
\phi_{R^{\prime\prime}L^{\prime\prime}\sigma}|\chi_{R^{\prime}L^{\prime}%
\sigma}^{\mathbf{k}}\rangle
\]
with
\[
\langle\phi_{RL\sigma}|\chi_{R^{\prime}L^{\prime}\sigma}^{\mathbf{k}}%
\rangle=\delta_{RR^{\prime}}\delta_{LL^{\prime}}+o_{RL}h_{RL,R^{\prime
}L^{\prime}}^{\mathbf{k}\sigma}.
\]
We have used $V_{RL,R^{\prime}L^{\prime}}^{\sigma}=V_{RL}^{\sigma}%
\delta_{RR^{\prime}}\delta_{LL^{\prime}}$, an assumption which is confirmed
numerically. Further, the diagonal overlap matrix $o$ as well as the
hamiltonian matrix $h$ are given in \cite{jepsen-TB}. Consider next
$V_{RL}^{\sigma}$. Assuming a spin-independent Hubbard $U$, and a diagonal
spin-density matrix $n_{RLL^{\prime}}^{\sigma}=n_{RL}^{\sigma}\delta
_{LL^{\prime}}$\cite{kk1} we obtain:
\begin{align}
V_{RL}^{\sigma}  &  =\sum_{L^{\prime}\sigma^{\prime}}U_{LL,L^{\prime}%
L^{\prime}}n_{RL^{\prime}}^{\sigma^{\prime}}-\sum_{L^{\prime}}U_{LL^{\prime
},L^{\prime}L}n_{RL^{\prime}}^{\sigma}\nonumber\\
&  =\sum_{L^{\prime}}U_{LL,L^{\prime}L^{\prime}}n_{RL^{\prime}}^{-\sigma
}+(U_{LL,L^{\prime}L^{\prime}}-U_{LL^{\prime},L^{\prime}L})n_{RL^{\prime}%
}^{\sigma}.
\end{align}
Now $U_{LL,L^{\prime}L^{\prime}}$ is substantial for all $LL^{\prime}$ in
contrast to $U_{LL^{\prime},L^{\prime}L}$ which is rather small, except when
$L=L^{\prime}$. We shall use $U_{LL,L^{\prime}L^{\prime}}\equiv U$ independent
of ${L,L^{\prime}}$ and $U_{LL^{\prime},L^{\prime}L}\equiv J$ for $L\neq
L^{\prime}$ which result in the simple form
\[
V_{RL}^{\sigma}=(U-J)[1/2-n_{RL}^{\sigma}],
\]
where the double counting term suggested in Ref. \cite{zaan} has been added.
For a fixed value of $U$ and $J$, the matrix-elements are evaluated and added
to the LMTO-Hamiltonian prior to diagonalization. The density-matrix
$n_{RL}^{\sigma}$ is updated every iteration using the eigenvectors as well as
the overlap matrix. The corresponding term,  which has to be added to the
total energy functional, is given by
\[
E^{U}-E_{dc}=\frac{(U-J)}{2}[N-\sum_{RL\sigma}n_{RL}^{\sigma}n_{RL}^{\sigma}]
\]
where $N=\sum_{RL\sigma}n_{RL}^{\sigma}$. It should be noted that already the
simple form of the non-local potential gives rise to upper and lower Hubbard
bands with an energy separation given by $(U-J)$.

\subsection{Selfconsistent LDA+U}

The cRPA method requires as input eigenfunctions and eigenvalues (fixed during
the calculation) and delivers as output the Hubbard $U$ matrix. On the other
hand, the LDA+$U$ method needs a $U$-matrix (fixed during the calculation) as
input and gives as output eigenfunctions and eigenvalues. The main point of
the present work is to merge these two schemes in a selfconsistent way.

We summarize the iterative steps:

1. Firstly we do a normal cRPA calculation \cite{ferdi1} in order to achieve
the initial Hubbard $U$ matrix (iteration one; matrix $U_{1}$) to be used in
the LDA+$U$ calculation. \newline\newline2. After the LDA+$U$ calculation has
converged we save the output LDA+$U$ eigenfunctions and eigenvalues and use
these to calculate $U$ within cRPA (Eqs. (\ref{P},\ref{Wr},\ref{U})), in order
to find the updated $U$-matrix for the next LDA+$U$ calculation (iteration
two; matrix $U_{2}$). \newline\newline3. The procedure is continued until the
$U$-matrix is stable i.e after $n$ iterations we have $U_{n+1} \approx U_{n}$.
\newline\newline

The size of the $U$-matrix is rather large, however many elements are related
by symmetry.

\section{Results and discussion}

All the results presented in this paper used the simple form of the non-local
potential, because a substantial number of tests have shown that more
elaborate forms of the potential do not influence the final results. The most
important finding in the present work is indeed the possibility to converge
the $U$-matrix within the defined self-consistency cycle. In all cases studied
convergency is reached within a reasonable number of iterations.

To illustrate the applicability of the present scheme to real materials we
have applied the scheme to NiO, which is an epitome of the charge transfer
insulators, and Gd. These two systems have been extensively studied both
experimentally and theoretically. The NiO LDA band gap is known to be too
small and likewise the LDA exchange splitting in $4f$ Gd is too small. These
provide a motivation for improving upon the LDA.

A summary of some results for our prototype systems: For NiO, the
self-consistent determined values $U=6.6$ eV and $J=0.9$ eV, improves the
bandgap (2.5 eV) , compared with conventional LDA, though too small in
comparison with experiment (4 eV). The exchange spin-splitting of the
$f$-bands in Gd are found to compare rather well with experiment ($\sim12-13$
eV) using our self-consistent determined values of $U=12.4$ eV and $J=1.0$ eV.
We have also calculated the Gd (NiO) magnetic moment to be $\mu=7.8 $ (1.5),
which is comparable to the experimental value $\mu=7.6 \;\mu_{B}$\cite{Gdmom}
(1.6-1.9 $\mu_{B}$), and an improvement compared to LDA.

\begin{figure}[ptb]
\begin{center}
\centerline{
\includegraphics[height=15cm,angle=0]{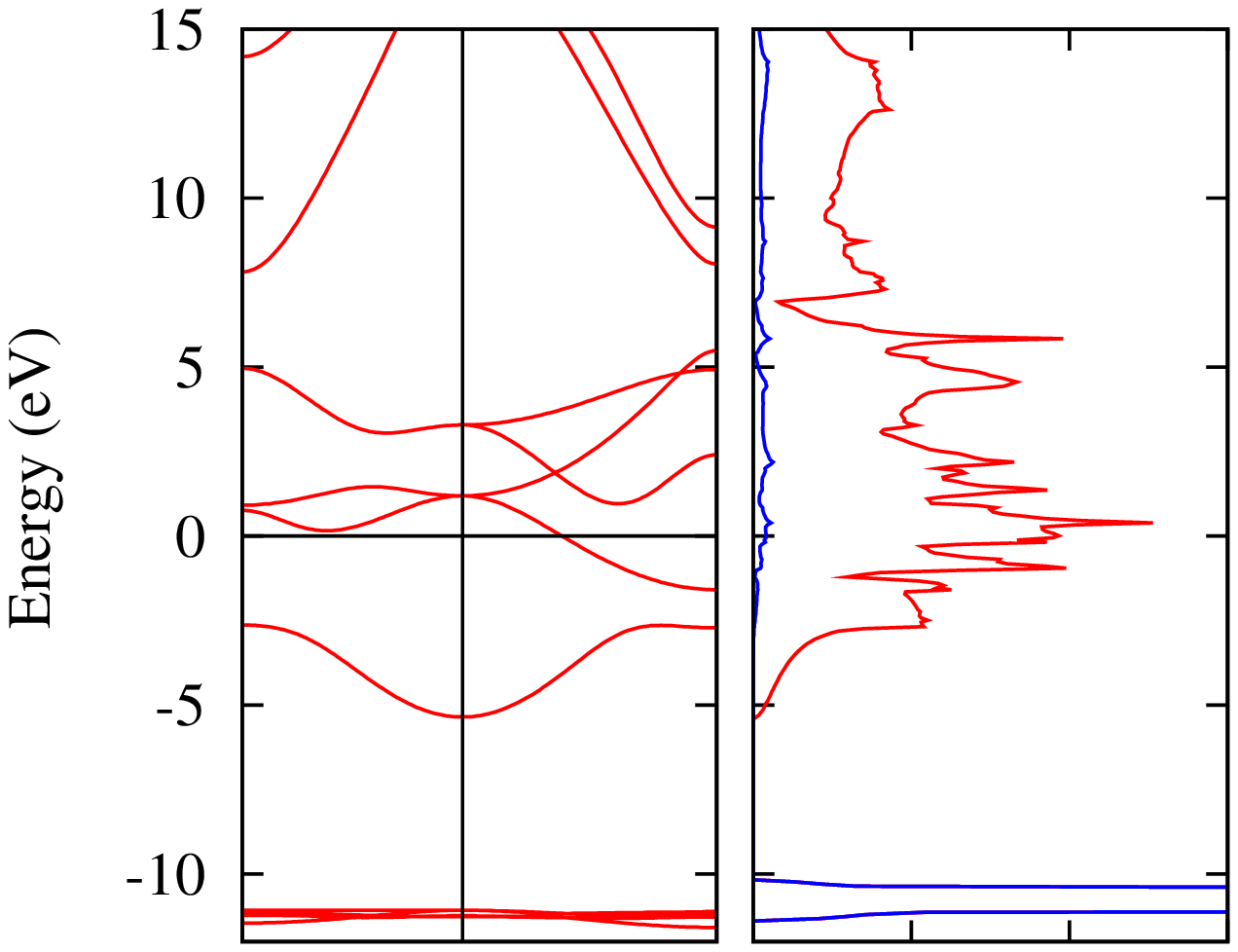}} \vspace{0.5cm}
\end{center}
\par
\hfill\caption{Gadolinium spin up bands using the \emph{self-consistent}
determined parameters: $U$ = 12.4 eV and $J$ = 1.0 eV. \newline Fermi energy
at 0 eV and the directions displayed are 1/2(1,1,1) $\rightarrow$ $\Gamma$
$\rightarrow$ (1, 0, 0). The corresponding total DOS and $f$ partial DOS are
also displayed. }%
\label{fig:Gd-up}%
\end{figure}

\begin{figure}[ptb]
\begin{center}
\centerline{
\includegraphics[height=15cm,angle=0]{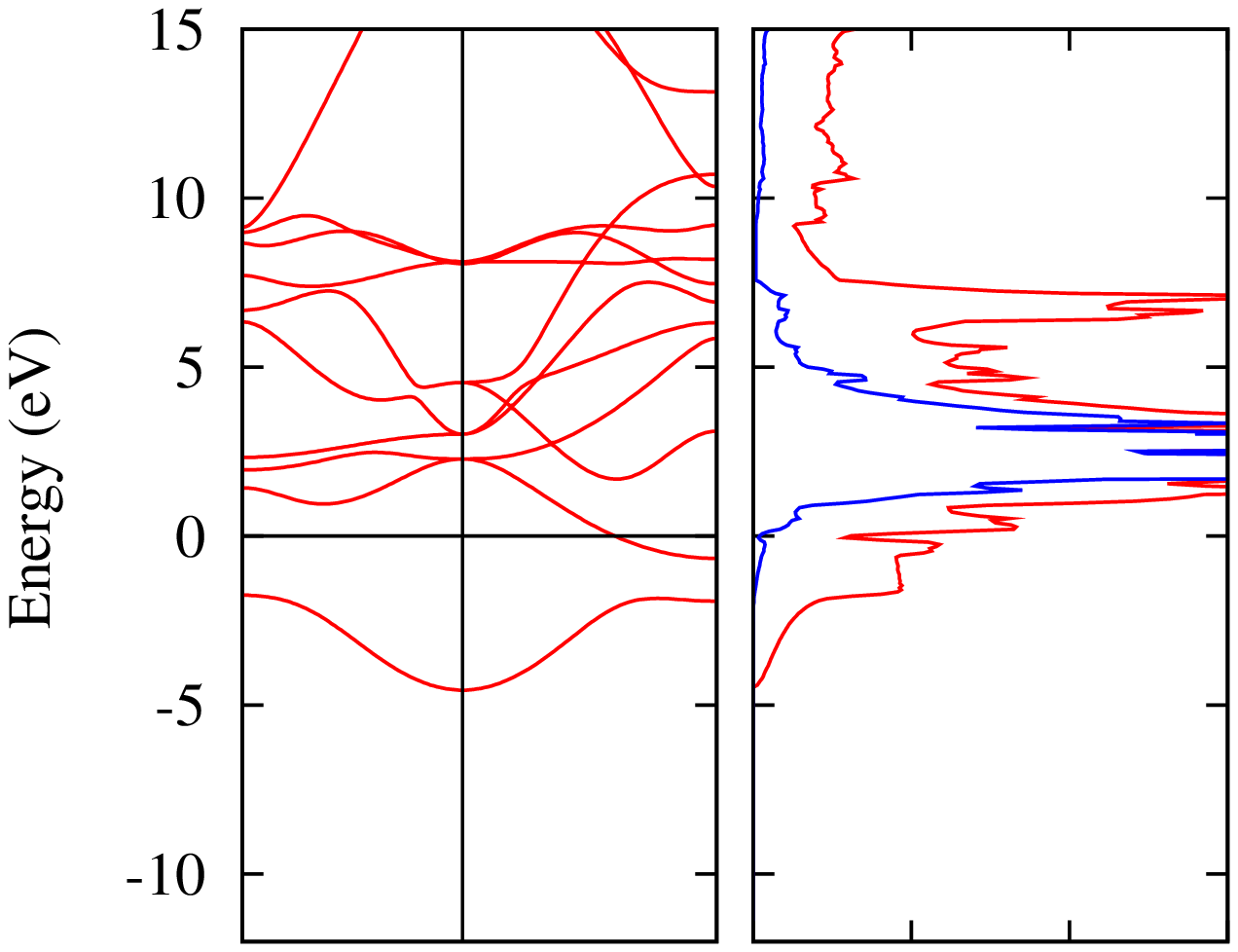}} \vspace{0.5cm}
\end{center}
\par
\hfill\caption{Gadolinium spin down bands using the \emph{self-consistent}
determined parameters: $U$ = 12.4 eV and $J$ = 1.0 eV. \newline Fermi energy
at 0 eV and the directions displayed are 1/2(1,1,1) $\rightarrow$ $\Gamma$
$\rightarrow$ (1, 0, 0). The corresponding total DOS and $f$ partial DOS are
also displayed.}%
\label{fig:Gd-do}%
\end{figure}

We first discuss Gd, where the LDA+$U$ bandstructure corresponding to the
self-consistent values of $U$ and $J$ are displayed in Figs. (\ref{fig:Gd-up}%
-\ref{fig:Gd-do}). The majority (spin up) $f$ bands are centred around -11 eV
and the minority ones around 3 eV. The occupied spin up bands are very narrow
due to shielding by the $5s$ and $5p$ electrons, due to the hybridization with
other bands the unoccupied minority bands display some dispersion, making it
difficult to extract the exchange splitting. However, we estimate that our
calculated exchange splitting at convergency is somewhat too large by say
$\sim1-2$ eV.

We note that our parameters differs significantly from those previously used
in literature. Harmon \emph{et al}\cite{harmon} found $U=6.7$ eV and $J=0.7$
eV using a supercell approach. The experimental gap (splitting between the PES
and BIS main-peaks) is given by $E_{g}=E_{N+1}+E_{N-1}-2E_{GS}$, which from
purely atomic considerations is predicted to be $U+6J$, using $N=7$ spin-up
electrons in the groundstate (GS). With the parameters of Harmon \emph{et
al.}\cite{harmon}, an underestimation is obtained, resulting in a splitting of
11 eV. As can be seen in Fig. 2, the 4f states no longer form a narrow
atomic-like band but hybridized with other states in the same energy range.
Thus, the atomic picture used to estimate the exchange splitting may not be
valid anymore.

\begin{figure}[ptb]
\begin{center}
\centerline{
\includegraphics[height=6cm]{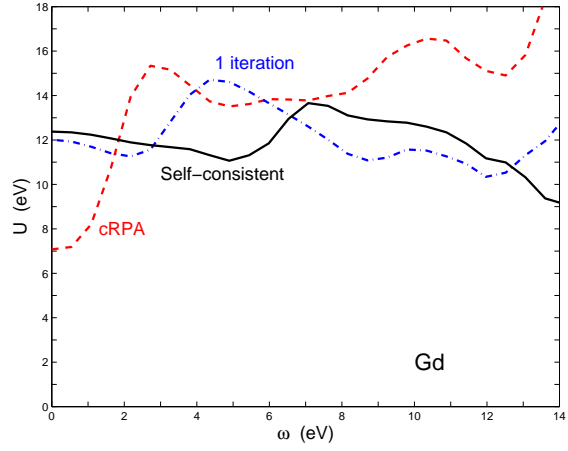}}
\end{center}   
\par
\hfill\caption{Frequency-dependent $U$ of gadolinium. }%
\label{fig:Uw-Gd}%
\end{figure}

The frequency-dependent $U$ \cite{kk2} from the normal cRPA calculation, i.e.,
from calculation starting from the LDA bandstructure, is shown in Fig.
(\ref{fig:Uw-Gd}). We note the dramatic change in $U$ for small energies,
shooting up to the self-consistent value of $U$ already within 2 eV. In fact,
the frequency dependence would have become even stronger if we had not used a
life-time broadening when calculating the response function. Using a
tetrahedral method for the Brillouin zone integration without a life-time
broadening would probably result in a decrease in $U$ from its zero-energy
value before it shoots up to a large value at around 1.5 eV. This behavior is
in contrast to the transition-metals studied earlier\cite{ferdi1}. Towards
self-consistency we noticed a significant change in $U$ already in the second
cRPA calculation (1 iteration) ; $U$ is in fact enhanced for small energies
giving rise to a quite smooth curve with weak dependency on frequency. As seen
in Fig. (\ref{fig:Uw-Gd}), the frequency dependence of $U$ is indeed much
weaker after self-consistency, with a relatively constant value of $U=12$ eV
in the frequency range around 5 eV. The weakening of the energy dependence of
\emph{U} for small energies may be explained by the increase in the exchange
splitting of the up and down $4f$ states. As the occupied $4f$ states are
pushed down the excitation energies from the occupied $4f$ states to
unoccupied states increase. Similarly, as the unoccupied $4f$ states are
pushed up, the excitation energies from occupied states to the unoccupied $4f$
states increase. Thus, the peak structure in the imaginary part of the
screened interaction arising from these excitations is shifted to higher
energy. Through the Kramers-Kronig relation this results in much smoother
behavior of \emph{U} at low energy. This result is very encouraging since it
gives justification for using a static value of \emph{U}.

\begin{figure}[ptb]
\begin{center}
\centerline{
\includegraphics[height=16cm]{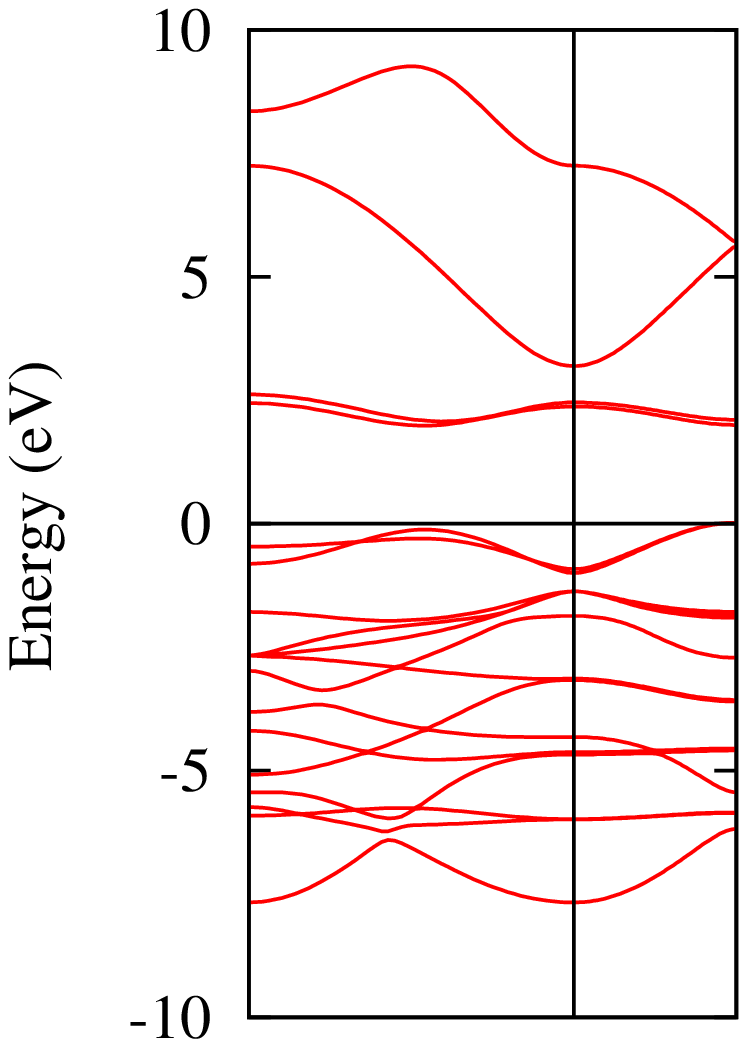}}
\end{center}
\par
\hfill\caption{NiO bands using the \emph{self-consistent} determined
parameters: $U$ = 6.6 eV and $J$ = 0.9 eV. Fermi energy at 0 eV and the
directions displayed are 1/2(1,1,-1) $\rightarrow$ $\Gamma$ $\rightarrow$
1/4(1, 1, 1).}%
\label{fig:NiO-band}%
\end{figure}

Finally we consider NiO, where the LDA+$U$ bandstructure corresponding to the
self-consistent values of $U$ and $J$ are shown in Fig. ( \ref{fig:NiO-band}).
For this system cLDA calculations yields $U=8$ eV and $J=1$ eV \cite{zaan},
which is comparable to our self-consistent values of $U$ = 6.6 eV and $J$ =
0.9 eV. The gap obtained using the cLDA parameters is 3 eV\cite{zaan},
compared to the experimenal gap of 4 eV\cite{fujimori}. The difference between
our and the cLDA $U$ (1.4 eV) is reflected in our decreased bandgap of 2.5 eV.

\vspace{2cm} \begin{table}[h]
\caption{A summary of results for $U$ and $J\,$ obtained with the present
method in comparison with other methods (in brackets).\ We compare also the
magnetic moments with experimental findings (in brackets).}%
\centering
\begin{tabular}
[c]{llll}\hline\hline
& $U$ (eV) & $J$ (eV) & Magnetic moment ($\mu_{\text{B}}$)\\\hline
NiO & 6.6 (8.0 \cite{zaan}) & 0.9 (1.0 \cite{zaan}) & 1.5 (1.6-1.9
\cite{alperin, fender})\\
Gd & 12.4 (6.7 \cite{harmon}) & 1.0 (0.7 \cite{harmon}) & 7.8 (7.6
\cite{Gdmom})\\\hline\hline
\end{tabular}
\label{tab:I}%
\end{table}

\vspace{2cm}

As in the case of Gd, the Hubbard \emph{U} as a function of frequency
undergoes a significant change as self-consistency is achieved. Starting from
the LDA bandstructure, the resulting \emph{U} calculated using the cRPA method
exhibits a strong energy dependence at low energy. As the band gap increases,
the energy dependence of \emph{U} at low energy becomes smoother. The
explanation of this behavior is similar to the case of Gd, namely, as the gap
increases the peak structure in the imaginary part of the screened interaction
is shifted to higher energy, and through the Kramers-Kronig relation, it
results in a smooth behavior of \emph{U} at low energy.

The too large 4$f$ separation obtained in the self-consistent LDA+$U$ scheme
may arise from a shortcoming of the LDA+\emph{U} scheme itself rather than in
the RPA used in calculating \emph{U}. A similar problem is also observed in
the so-called Quasiparticle Self-consistent GW (QSGW) scheme \cite{chantis}.
As the 4$f$ separation becomes larger, the screening associated with the 4$f$
bands becomes weaker, which leads to a larger $U$. This in turns tends to
decrease the screening strength and so forth. This indicates a shortcoming of
the theory, namely, the absence of energy-dependent self-energy and the vertex
correction, the latter is also left out in the QSGW schemes.

A further problem that plagues the LDA+$U$ scheme is the double-counting
problem. This problem becomes apparent when the relative position of the
correlated bands with respect to other bands is important. This relative
position is rather sensitive to the double-counting formula used in the
scheme. We believe this double-counting problem is responsible for the
incorrect positioning of the 4$f$ bands in Ce as well as the 3$d$ bands in
NiO, giving a too small band gap in the latter, which has also been found in
other works \cite{blaha}. While the separation between the unoccupied $e_{g}$
and occupied $t_{2g}$ bands of nickel is reasonably well reproduced, the
relative position of these 3$d$ bands with respect to the oxygen 2$p$ bands is
presumably incorrect. In the case of the transition metal oxides, such as NiO,
the band gap is formed between the unoccopied Ni $e_{g}$ band and the occupied
O 2$p$ band.

\section{Conclusion}

We have developed a new self-consistent LDA+$U$ scheme, in which the important
parameter \emph{U} is determined self-consistency using the cRPA method. As
test cases we have considered NiO and Gd and it is shown that the scheme does
yield converged results. The exchange splitting in Gd has been found to be
too
large by 1-2 eV whereas the band gap in NiO has been found to be too small,
2.5 eV compared with the experimental value of about 4.0 eV. An interesting
finding is that the energy dependence of \emph{U} at low energy is found to be
much smoother after self-consistency compared with the result obtained from
the LDA bandstructure. This provides justification for using a static value of
\emph{U}. Our results indicate some short-comings of the LDA+\emph{U} scheme,
in particular the incorrect positioning of the 4f states in Gd and the 3d
states in NiO points to a problem with the double-counting term. Investigating
different forms of the double-counting term within the newly developed
self-consistent LDA+\emph{U} scheme could be a fruitful direction to pursue in
the future.

\section{Acknowledgment}

We greatly acknowledge discussions with Olle Gunnarsson and Janusz Kanski and
FA acknowledges support from the G-COE program of MEXT (G-03).

\newpage

\noindent

\noindent

\noindent


\begin{thebibliography}{99}
\bibitem {kk1}For all the materials studied, the off-diagonal elements of the
spin-density matrix have been very small.
\end{thebibliography}
\end{document}